\documentclass[twocolumn]{aastex63}
\usepackage{natbib,amsmath,mathtools,float}
\usepackage{color,ulem,epstopdf}
\usepackage{graphicx}
\usepackage{graphics}
\usepackage{verbatim}

\usepackage{CJKutf8}

\begin{document}

\begin{CJK*}{UTF8}{gbsn}
\bibliographystyle{apj}

\shorttitle{New Population -- sub-Earths }
\shortauthors{Qian \& Wu}

\title{A Distinct Population of Small Planets: Sub-Earths}

\author{Yansong~Qian (钱岩松)}

\email{yansong.qian@mail.utoronto.ca}

\author{Yanqin~Wu (武延庆)}
\affil{Department of Astronomy and Astrophysics, University of Toronto, Toronto, ON M5S 3H4, Canada}
\email{wu@astro.utoronto.ca}

\begin{abstract} 

The sizes of small planets have been known to be bi-modal, with a gap separating planets that have lost their primordial atmospheres (super-Earths), and the ones that retain them (mini-Neptunes). Here, we report evidences for another distinct population at smaller sizes. By focussing on planets orbiting around GK-dwarfs inward of 16 days, and correcting for observational completeness, we find that the number of super-Earths  peak around $1.4$ Earth radii and disappear shortly below this size. Instead, a new population of planets (sub-Earths) appear to dominate at sizes below $\sim 1$ Earth radius, with an occurrence that rises with decreasing size.  This pattern is also observed in ultra-short-period planets.

The end of super-Earths supports earlier claims that super-Earths and mini-Neptunes, planets that likely form in gaseous proto-planetary disks, have a narrow mass distribution. The sub-Earths, in contrast, can be described by a power-law mass distribution and may be explained by the theory of terrestrial planet formation. We therefore speculate that they are formed well after the gaseous disks have dissipated. 
The extension of these sub-Earths towards longer orbital periods, currently invisible, may be the true terrestrial analogues. This strongly motivates new searches.
\end{abstract}

\keywords{planets}

\section{Introduction}

Most extra-solar planets known to date are likely formed in gaseous protoplanetary disks. These include the Jovian planets, the super-Earths and their siblings, the mini-Neptunes -- the latter two  are sometimes jointly called 'Kepler planets' for short  \citep[e.g.,][]{Borucki2010,Lopez,WuLithwick,Fulton}. However, there is little evidence for the existence of terrestrial planets outside our own solar system. Terrestrial planets are thought to have formed from collisional debris, well after the dispersion of proto-planetary disks. We term these Generation II planets, to distinguish them conceptually from the first group (Generation-I planets).

Currently, evidences for Gen-II planets are best seeked using data from the {\it Kepler} mission. The final {\it Kepler} mission transit planet search \citep[DR25,][]{Twicken} includes nearly 200,000 stellar targets  and yields $\sim 4700$ planet candidates.
Such a large sample allows us to examine the exoplanet populations in detail.  A majority of these candidates are super-Earths and  mini-Neptunes. It has been convincingly shown that these two populations are separated by a radius gap \citep{Fulton,vaneylen}, readily explained by the theory of photo-evaporation \citep{Lopez,OwenWu13,OwenWu17,Jin2014}, or theory of core-powered mass-loss \citep{hilke}. As such, the two populations have the same origin, with super-Earths being former mini-Neptunes that happen to lie too close to their host stars. Moreover, the sizes of these planets appear to be narrowly distributed\footnote{This is already  clear for the mini-Neptunes, and becomes clear for the super-Earths as well in this work.},
and rise with the masses of the host stars \citep{cks7,Wu2019,Cloutier2020}.
 
 On the other hand, the initial aim of {\it Kepler}, Earth-like planets, remain elusive. While there are a smattering of small planets at short orbital periods that are Earth-sized, {\it Kepler} has a very low survey completeness for such planets at AU-distances. It is also unclear if the close-in ones are low-mass extension of the super-Earths (and are therefore Gen-I planets), or are a distinct population. This is the very question we set out to answer in this work.

 Our effort is made possible thanks to two recent advances. The planet detection efficiency of the DR25 pipeline  has now been characterized for every star by injection tests at the flux-level \citep{burkeFLIT}, and characterized for the group as a whole by injection tests at the pixel-level \citep{christainsenpl}.
 Meanwhile, uncertainties in the radii of KOIs  have now been dramatically reduced using GAIA DR2 stellar radii \citep{Berger2020pc}. 

In this work, we focus on planet candidates that orbit around GK dwarfs, with periods shortwards of $16$ days. We are able to show that the super-Earths, like the mini-Neptunes, have a narrow size distribution. Their occurrences fall off rapidly below a peak. This feature enables us to detect the rise of a new population at smaller sizes, the sub-Earths. The size distribution exhibits clearly an inflection point 
between super-Earths and sub-Earths, most likely a gap.

Results similar to ours, that there is a new population of small planets, has been obtained or alluded to previously.
\citet{NeilRogers} found that {\it Kepler} data is best explained by three populations of exoplanets:
gaseous (mini-Neptunes), evaporated cores (super-Earths) and intrinsically-naked. However, their joint distribution  does not show an inflection point, nor a gap, as we find here.
In a different approach, \citet{hsu2019} studied the planet occurrence rates across a wide range of orbital periods and planet radii. Their radius bins are broader than ours here, so any possible gap feature is somewhat smeared. They do, however, report an excess of small planets, consistent with our results here.
 
Both these works aim to determine the more general landscape of small planets, while ours drills down to one specific question. This allows us to focus on the most revealing parameter space and to slim down the model. We are also able to test the robustness of our results by experimenting with different sample selections and completeness functions.

\section{Planet and Star Samples}

Starting from the Kepler DR25 catalogue of $200,038$ host stars and 4034 planet candidates \citep{dr25Thompson} (excluding false positives, NASA Exoplanet Archive), we introduce a number of cuts to obtain a suitable sample for this study. We give our justifications for these cuts and list the resulting sample sizes in Table \ref{table:cut}. In \S \ref{subsec:cuts}, we explore how results may be impacted by different sample cuts.

\begin{enumerate}

\item {\bf GAIA-able:} We only consider stars that are in the Gaia-Kepler Stellar (GKS) Catalog  \citep{Berger2020star}. In addition, we follow \citet{Berger2020pc} to exclude stars with Gaia DR2 re-normalized unit-weight error (RUWE) larger than 1.2 and with isochrone-derived  goodness-of-fit parameters lower than 0.99. These cuts can remove binary stars and ensure the reliability of parameters. Compared to the much larger radius uncertainty in the KIC catalogue \citep[typically $25\%$,][]{Fulton}, the GKS catalogue employs GAIA parallaxes to achieve a much lower uncertainty of $\sim 4\%$ in stellar radius \citep{Berger2020star}. This allows us to separate planets into smaller radius bins and to observe the fine details in planet occurrence. 

\item {\bf GK-dwarf:}
It has been noticed before \citep{Fulton,Wu2019,Berger2020pc,Cloutier2020} that the sizes of both super-Earths and mini-Neptunes rise systematically with stellar mass. Such a trend is as yet unexplained. For our purpose, this tends to smear any features in the planet size distribution. This leads us to focus only on  GK-dwarfs, within a stellar radius range of $R_* \in [0.75R_\odot,1R_\odot]$ (and a similar stellar mass range). It is beneficial to study smaller stars, as small planets are more readily detectable around them. This choice of GK-dwarfs also ensures a sufficiently large sample. 
Finally, we choose our sample based on stellar radius, rather than stellar mass as provided by \citet{Berger2020star}. The former has a median uncertainty of $4\%$, while the latter is larger at $7\%$.

\item {\bf Brightness:} We discard  stars with $K_p$ magnitudes  above 15.  Very few Earth-sized planets are discovered around dim stars, and both the false-positive probabilities and radius errors rise for planet candidates around these stars \citep{Petigura2013}.

\item {\bf NoisyTargetList} is a list of Kepler stars with too high noise to apply the the detection efficiency model \citep{Burkemodel}.
Thus, we exclude stars on this list.

\item {\bf False Positives:} we remove  planet candidates with a false-positive-probability $FPP > 0.1$, based on the False Positive Probabilities(FPP) Table produced by \citet{Morton2016}. This alleviates confusion by eclipsing binaries. For the remaining ones, we assume a reliability of $100\%$ (more on this below).

\item {\bf Planet Radius}: we remove large planets with radii above $4 R_\oplus$ and small planets below $0.6 R_\oplus$. Below $0.6 R_\oplus$, the detection efficiency drops too low to allow any secure statement.

\item {\bf Period range:} a major distinction in this study is our focus on short period planets, inward of $16$ days. By zooming in to the region where small planets can potentially be discovered (detection efficiency above $\sim 10\%$), we contain the uncertainties inflated by rare detections. While this reduces our planet sample, it actually helps increasing our sensitivity to small planets.

\end{enumerate}

Eventually, we are left with  $298$ planet candidates and $13,297$ qualified stars.  In the following, we consider the pipeline completeness for this sample.

\begin{table}[H]
\centering
\begin{tabular}{p{0.35\linewidth}p{0.23\linewidth}p{0.23\linewidth}}

\hline
Cut & $n_p$ & $N_*$\\
\hline
Exoplanet Archive & $4034$ & $200,038$\\
GAIA-able &   $3270$ & $162,657$ \\
$0.75 < R_*/R_\odot < 1$ & $1112$ & $34,399$ \\
$K_p \leq 15$ & $590$ & $13,498$ \\
NoisyTargetList & $580$ & $13,297$\\
FPP $<0.1$ & $505$ & - \\
$0.6 \leq R_p \leq 4 R_\oplus$ & $469$ & - \\
$1\leq P_{\rm orb} \leq 16$ &$298$ & -\\
\hline
\end{tabular}
\caption{Star and Planet Sample Cuts.}
\label{table:cut}
\end{table}

\section{Completeness}

For every planet in the {\it Kepler} field, its detection depends on a number of factors: geometric probability of transit; pipeline efficiency in identifying the transit as a threshold-crossing-event (TCE); and loss during the vetting process. In the following, we discuss the last two factors, together with the issue of reliability. 

\label{sec:completeness}
\subsection{Pipeline Detection Efficiency}

By injecting simulated transit signals into real light curves and then recovering them using the same search pipeline,  one can determine the so-called 'pipeline efficiency' as a function  of planet size, orbital period and stellar properties. The {\it Kepler} team have published multiple studies of pipeline completeness \citep[e.g.][]{Christiansen2016, Thompson2018}. The most recent ones are those by \citet{burkeFLIT,christainsenpl} for the DR25 pipeline. The \citet{christainsenpl} work is based on the so-called pixel-level transit injection tests. Limited by the computation capacities, each {\rm Kepler} star only receives one planet injection and the resulting completeness is an average over the entire star sample. In contrast, the much simpler flux-level tests \citep{burkeFLIT} can inject many different signals into a given star, and one can survey a range of planet parameters (radius and period) on a per-star basis, to obtain
the detection efficiency, $DE(p,r,\theta_*)$, as a function of orbital period, planet radius and host star properties.  Such a model captures the target-to-target variations and is recommended for studies involving a  small subset of {\it Kepler} targets, as is the case here.
We will adopt this model and use the KeplerPorts python code \citep{Burkemodel} to generate the pipeline efficiency.

In detail, KeplerPorts calculate the detection efficiency, $DE(P,R_p,\theta_*)$, based on the so-called multiple event statistics,  $MES(P,R_p,\bar{\theta})$ \citep{Jenkins,Christiansen12}, with $\bar{\theta}$ being stellar parameters ($R_*$, $\log g$ and $T_{\rm eff}$). This statistics is closely related to the transit signal-to-noise ratio. The detection efficiency 
is also empirically found to correlate strongly with the so-called CDPP slope, the temporal behaviour of the stellar noise. This dependency is taken into account by the \citet{Burkemodel} model,
based on injection experiments. Furthermore, the method of 'MES Smearing' is applied to transits with non-zero impact parameters. 

\begin{figure*}
\includegraphics[width=0.48\textwidth,trim=0 0 0 0,clip=]{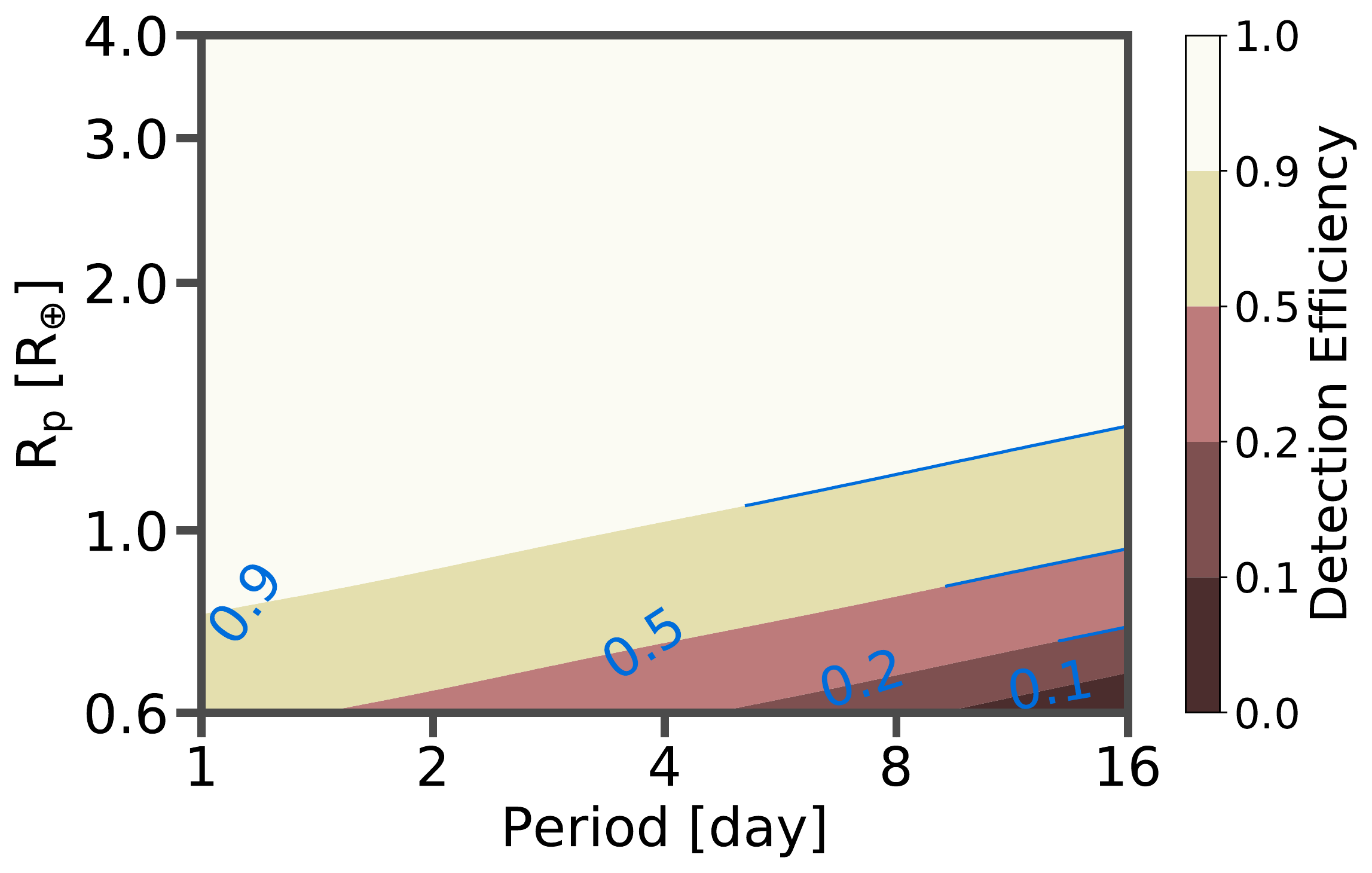}
\includegraphics[width=0.48\textwidth,trim=0 0 0 0,clip=]{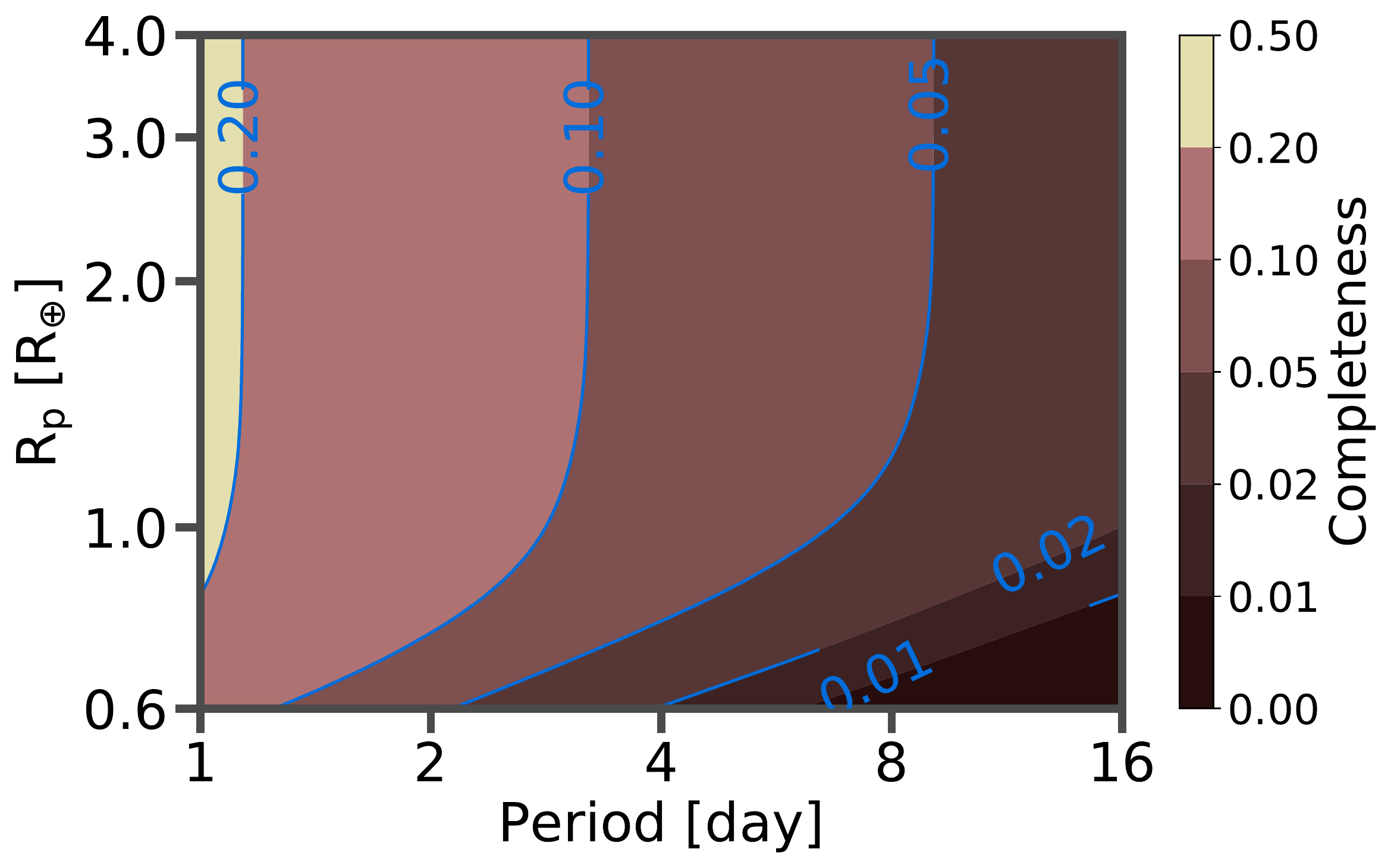}
\caption{ Detection completeness. The left panel shows the  averaged pipeline detection efficiency for the GK-dwarfs in our sample, as a function of planet radius and period, obtained using the \citet{burkeFLIT} model. Within 16 days, nearly all transiting planets with radii greater than $1.3 R_\oplus$ are recovered.
Moreover, at least half of the planets with radii greater than $0.9 R_\oplus$ are recovered.
The right  panel shows the overall  efficiency (eq. \ref{eq:eta}) when transit probability is also considered.}
\label{fig:DEaverage}
\end{figure*}

The resultant detection efficiencies are then averaged over the $N_* =  13,297$ GK-dwarfs in our sample. For the radius and period ranges of interests, these efficiencies are plotted in the left-hand panel of Fig. \ref{fig:DEaverage}. One observes that, inward of $16$ days, transiting planets larger than $1.3 R_\oplus$ can  be detected fairly completely  ($DE \geq 90\%$), and the efficiency drops below $\sim 50\%$ for planets smaller than $0.9 R_\oplus$. The right-hand panel presents the overall efficiency, $\eta (p, r)$, one that also includes the geometric transit probability.
 This is  calculated as 
 \ref{fig:DEaverage}.
\begin{equation}
    {\eta}(P,R_p) =\frac{1}{N_*} \sum_{i=1}^{N_*}DE_i(P,R_p) \times P_{\rm geo, i}(P,R_p)\, ,
    \label{eq:eta}
\end{equation}
where $P_{\rm geo,i} = 2 R_{*,i}/a$ and we adopt an estimate for the star mass from \citet{Berger2020star}. 

\subsection{Vetting Completeness}

\begin{figure}
\includegraphics[width=0.48\textwidth,trim=0 0 0 0,clip=]{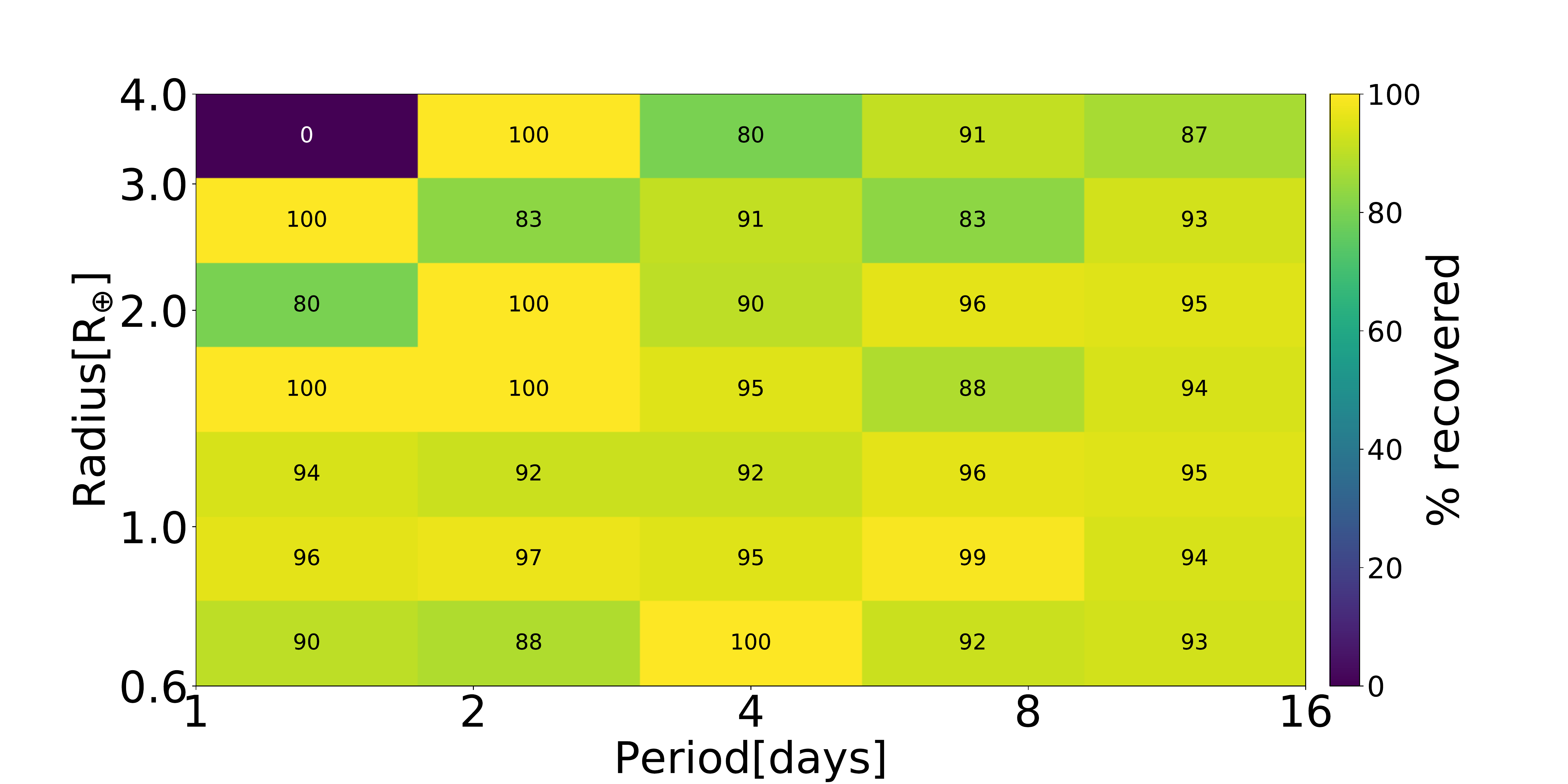}
\caption{Vetting efficiency for our ranges of planet radius and orbital period, obtained using Robovetter by \citep{Coughlin}. On average, $93\%$ simulated planets within this range pass the vetting tests.}
\label{fig:vet}
\end{figure}

Binary stars and some astrophysical noise can masquerade as transit signals. So every TCE needs to be vetted before qualifying as a planet candidate. This can introduce a reduction in completeness as some genuine planets may be vetted out as false positives. This vetting efficiency is presented in Fig \ref{fig:vet} for our ranges of interest, produced using  the robotic vetting code Robovetter \citep{Thompson2018,Coughlin}. For the short period planets we are interested in, the vetting efficiency 
has an average of $93\%$, with no obvious dependency on planet radius or period. So in this study, we simply assume this to be $100\%$ for the entire space.

\subsection{Reliability}

\citet{Morton2016} assigned  a false positive score (FPP) for each planet candidate, using an automated validation procedure. This uses a full range of observational constraints to calculate the probabilities of various hypothesis scenario (genuine planets, brown dwarfs, eclipsing binaries, etc). We have discarded all False Positives, but retained candidates in both the 'confirmed' and the 'candidate' categories,  with the proviso that ${\rm FPP} < 0.1$. We follow convention and define a reliability index $=(1 - {\rm FPP})$. These are shown in Fig. \ref{fig:samples} by the colors of the dots. Since reliability is high in our sample, we simply ignore this distinction and assume all candidates are genuine.

\section{Size Distribution of Small Planets -- A New Gap}

\begin{figure*}
\includegraphics[width=0.85\textwidth,trim=0 0 0 0,clip=]{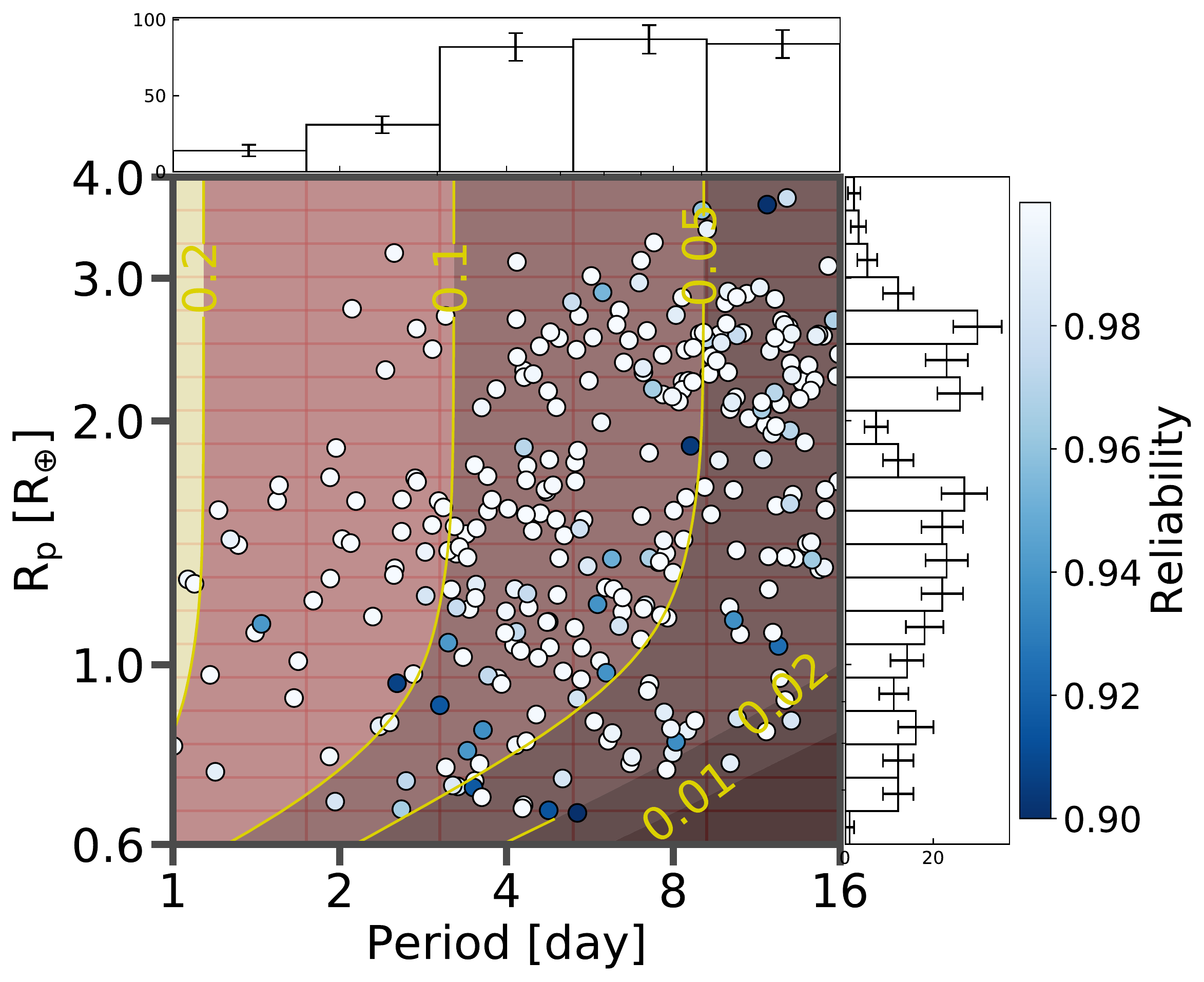}
\caption{ The sample of $280$ planet candidates in our analysis, plotted as their observed radii vs. periods. The contours indicate the total detection efficiency (including transit probability), and the color of each dot indicates the reliability index (1-FPP).  The size histogram and period histogram are shown to the left and the top of the main figure. Even without correcting for completeness, one can already see the two gaps, one between super-Earths and mini-Neptunes at around $2R_\oplus$, and one between sub-Earths and super-Earths at around $1 R_\oplus$.
}
\label{fig:samples}
\end{figure*}

Equipped with the planet sample and their detection efficiencies (Fig. \ref{fig:samples}), we proceed to recover the underlying distribution. We first present our analysis using the straight-forward IDEM method, before elaborating on a Bayesian approach. The two are shown to produce compatible results.

In the following, we divide our planet sample on a logarithmic grid of period and radius. To ensure a sufficient radius resolution, we choose to use  $20$ radius bins from $0.6$ to $4R_\oplus$. To ensure a sufficient number of candidates in each bin, we choose a relatively crude period resolution, $5$ bins from $1$ to $16$ days. Even with this choice, there are some bins that contain zero or only one candidate and their information contents have to be carefully evaluated (see below). 

\subsection{IDEM: inverse detection efficiency method}

One straight-forward, non-parametric way of obtaining the occurrence rate is called 'IDEM' \citep{Howard2012,Petigura2013}. Assuming the observed population is drawn from an underlying population weighted by the detection efficiency, one can retrieve the differential occurrence rate  by dividing the number of planets detected in the bin $\Delta \ln P \Delta \ln R_P$,  $n(\rm P,R_P)$, by the detection efficiency,
\begin{equation}
{\frac{d^2 f }{d \ln P d\ln  R_P}}  = {\frac{1}{\Delta \ln P \Delta \ln R_P}} \times {\frac{1}{N_*}}\, \times {\frac{n(\rm P, R_P)}{ \eta(P, R_P)}}\, ,
\label{eq:idem}
\end{equation}
where $\eta(P, R)$ is the detection efficiency, suitably averaged over the relevant grid, and $N_*$ is the stellar sample size. We then sum over the period bins to obtain a size distribution 
\begin{equation}
    {{df}\over{d\ln  R_P}} = \sum {{d^2 f}\over{d\ln P d\ln R_P}} \Delta \ln P\, .
\end{equation}

The dominant source of error is Poisson error, as opposed to measurement errors. To obtain an estimate on the uncertainty, we note that since we  are dealing with small candidate numbers in each bin (some bins even contain  zero detection),  the confidence interval cannot be estimated using the usual $1/\sqrt{n}$ approach. Instead, 
we adopt Table VI of \citet{Feldman1998} to obtain the $95\%$ confidence limits when presented with  a detection $n(P,R_P) \geq 0$. When summing over different period bins, we assume that the confidence limits from each bin are independent and adds the  errors quadratically. The uncertainty range will then rise with bin number. This, we believe, yields a maximal uncertainty range. 

This approach departs from previous IDEM studies \citep[e.g.][]{Howard2012,Petigura2013,kunimotomattews} in providing  a more conservative error estimate.
To obtain  a more realistic and smaller uncertainties, one  needs to adopt a
 model-based approach (e.g., Bayesian, see below) that accounts for correlated information among bins, while paying a price for extra assumptions.

The result of this analysis is presented in tandem with the Bayesian model in Fig. \ref{fig:idembay}.

\subsection{Bayesian Modelling}
\label{sec:models}

\subsubsection{Model Ansatz}

To conduct a Bayesian analysis, we need to make an array of assumptions for the underlying distribution. 

First, we will depart from tradition and consider the distribution in planet mass and period. We use planet core-radius ($R_c$) as a proxy for its mass \citep[roughly, $M_p \propto R_c^4$,][]{Fortney2007}.  Since mini-Neptunes and super-Earths are of the same origin and therefore likely share the same mass distribution, it seems sensible to combine their signals together by working on planet mass. Moreover, this enables us to separate the distribution in  period-radius space (see below).

While $R_c \approx R_P$ for super-Earths, mini-Neptunes need some explanation. According to the theory of photo-evaporation, the radii of the two populations are spaced by a ratio that is nearly $2$. We adopt a number $1.7$ taken from \citet{OwenWu17,Wu2019}. This ratio is the result of fitting the observed distributions with evaporation theory. It is also the ratio between the observed super-Earth peak ($\sim 1.4 R_\oplus$) and mini-Neptune peak ($\sim 2.4 R_\oplus$) in our sample.  So in our study, for any planets lying above $2 R_\oplus$ (the so-called 'Fulton'-gap), we simply divide their observed radii by $1.7$ to obtain their core radii. 
When correcting for detection completeness, we note that while super-Earths are smaller and are in principle more incomplete compared to mini-Neptunes, the difference is minor  and can be ignored (Fig. \ref{fig:samples}). 

One naturally wonders if this procedure produces artificial features. We 
show that folding Neptunes into Super-Earths, while enhancing the super-Earth peak, is not critical to our conclusions on the existence and characters of the sub-Earths.

Next, we assume that the planet distribution in the $P-R_c$ plane is separable, 
\begin{equation}
{{d^2 f}\over{d \ln P\, d \ln R_c}}  = {{d f_P}\over{d\ln P}} \times {{df_R}\over{d\ln R_c}}\, .
\label{eq:separation}
\end{equation}
Such a separability assumption, often adopted in literature \citep[e.g.][]{Silburt,Mulders2018,NeilRogers}, is problematic when describing the distribution of super-Earths and/or mini-Neptunes individually, since their relative proportion depends on distances from the star.
It is, however, reasonable when describing the joint distribution. Studies of the planet mass distribution \citep[e.g.][]{Wu2019} have shown that it is relatively distance dependent. This is another reason behind our adoption of $R_c$ as the relevant parameter. 

In the following, we will experiment with different forms of $f_R$, while using the same broken-power-law for $f_P$,
\begin{equation}
    \frac{df_P}{d\ln P}  = 
    \left\{
    \begin{array}{@{}lr@{}}
    \left({{P}\over{\rm 1day}}\right)^\alpha, & \text{if}\    P\leq P_b \\
\left({{P_b}\over{\rm 1day}}\right)^{\alpha-\beta}\left({P\over{\rm 1day}}\right)^\beta, & \text{otherwise}
\end{array}\right.
        \label{eq:fP}
\end{equation}
where $P_b, \alpha, \beta$ are free parameters to be determined from data. This form is motivated by previous studies of super-Earths/mini-Neptunes \citep[e.g.][]{Silburt,Mulders2018,NeilRogers}. We assume that any new population of planets also satisfy the same rules.
We relax this assumption in  Model 4. 

For the planet mass (or core-radius) distributions, we adopt the following three models, each with a different physical significance.

 {\bf Model 1: one Gaussian.}
Here, we assume that nature produces only one single population of planets, for core-radius from  $0.6\oplus$ to $2\oplus$ (or from $\sim 0.2$ to $16 M_\oplus$, if these planets all share the terrestrial composition). We describe this using a log-Normal distribution in core-radius,
\begin{eqnarray}
    \frac{df_R}{d\ln R_{c}} &=&
      A\times  N(
      \ln R_c, \mu_0, \sigma_0) \nonumber \\
& = &   A \times  \frac{1}{\sqrt{2\pi \sigma_0^2}}\exp\left[-\frac{(\ln R_{c}-\ln \mu_0)^2}{2\sigma_0^2}\right]\, .
    \label{1gauss}
\end{eqnarray}
We adopt broad linear priors for 
for both $\mu_0$ and $\sigma_0$: 
$\mu_0 \in [0.6, 2] R_\oplus$  and  $\sigma_0 \in [0,1]$. 

{\bf Model 2: two Gaussians.} 
To allow for a separate population of planets, we assume the core-radius distribution is a superposition of two Gaussians, 
\begin{equation}
    {df_R\over {d\ln R_C}}  = B\times N(\ln R_c, \mu_1, \sigma_1) +C\times N(\ln R_c, \mu_2, \sigma_2)\, ,
    \label{2gauss}
\end{equation}
where the first term describes the Super-Earth/Neptune population, 
and we adopt linear priors of $\mu_1 \in [1.3,1.6]R_\oplus$ and $\sigma_1\in [0,0.5]$. These narrower priors are intuited from both the above IDEM analysis and previous studies \citep[e.g.][]{hsu2019,NeilRogers}.  For the second term, employed to describe a new lower-mass population , we assign broad linear priors of $\mu_2 \in [0.6, 1.2] R_\oplus$ and $\sigma_2 \in [0,1]$.

{\bf Model 3: power-law+Gaussian.} While super-Earth/Neptunes appear to be well described by (narrow) Gaussians, this does not appear so for the new population of planets we are after. The above 2-Gaussian results, as well our IDEM analysis, suggest that they may have a broad radius range, compatible with a power-law distribution. We  attempt the following power-law/Gaussian combination,
\begin{equation}
    {{df_R}\over{d\ln R_C}} = B\times N(\ln R_c, \mu_1, \sigma_1)+ D\times \left({{R_C}\over{R_{\oplus}}}\right)^\gamma\, .
    \label{powergauss}
\end{equation}
We  select a linear prior of $\gamma \in [-3,1]$. This is not restrictive, as it allows for both ascending and descending trends towards small sizes.
Moreover, neither this model nor the 2-Gaussian model presumes the presence of a gap-like feature.

{\bf Model 4: 2-periods.} The previous three models adopt the same period distribution for any planet population. 
Here we relax this assumption by allowing for a separate period distribution for each population (based on model 3),
\begin{eqnarray}
    {{d^2 f}\over{d\ln P d\ln R_C}} & = & B\times N(\ln R_c, \mu_1, \sigma_1)\times \left(\frac{df_P}{d\ln P}\right)_I\nonumber \\
    & & + D\times \left(\frac{R_{c}}{R_{\oplus}}\right)^{\gamma}\times \left(\frac{df_P}{d\ln P}\right) \, .
    \label{eq:2period}
\end{eqnarray}
Here  both period distributions have the same form as eq. (\ref{eq:fP}) but they can have different parameters. Moreover, 
since the super-Earths/Neptunes extend to well beyond $16$ days and their period-distribution has been well determined \citep[see, e.g.,][]{NeilRogers}, we can adopt narrower priors on their parameters based on previous studies (see Table \ref{table:parameter}).

\subsubsection{Bayesian Operations}

Here, we apply Bayesian inference to determine the parameters of maximum posterior probability.

\begin{table*}[ht]

\centering
\begin{tabular}{p{0.15\linewidth}p{0.15\linewidth}p{0.15\linewidth}p{0.15\linewidth}p{0.15\linewidth}p{0.15\linewidth}p{0.15\linewidth}}
\hline
\textbf{Parameter} & \textbf{1 Gaussian} & \textbf{2 Gaussian} & \textbf{Power-Gaussian} &
\textbf{2 Period} &
\textbf{Prior}\\
\hline
\hline
A & $0.008^{+0.003}_{-0.002}$ & - &-&-&$[0,0.05]$\\

$\mu_0$ & $1.31^{+0.13}_{-0.07}$ & - & - &-&$[0.6,2]$\\

$\sigma_0$ & $0.50^{+0.17}_{-0.08}$ & - & - &-&$[0,1]$\\
\hline

B & - & $0.0029^{+0.0013}_{-0.0010}$ &$0.0027^{+0.0010}_{-0.0008}$&$0.0031^{+0.0014}_{-0.0012}$&$[0,0.01]$\\

$\mu_1$ & - & $1.44^{+0.02}_{-0.03}$ & $1.43^{+0.02}_{-0.02}$ &$1.43^{+0.02}_{-0.02}$&$[1.3,1.6]$\\

$\sigma_1$ & - & $0.14^{+0.03}_{-0.02}$ & $0.14^{+0.02}_{-0.02}$ &$0.13^{+0.02}_{-0.02}$&$[0,0.5]$\\
\hline

C & - & $0.005^{+0.003}_{-0.003}$ &-&-&$[0,0.01]$\\

$\mu_2$ & - & $0.74^{+0.12}_{-0.09}$ &-&-&$[0.6,1.2]$\\

$\sigma_2$ & - & $0.6^{+0.3}_{-0.3}$ &-&-&$[0,1]$\\
\hline

D & - & - & $0.0028^{+0.0010}_{-0.0008}$ &$0.0030^{+0.0015}_{-0.0009}$&$[0,0.01]$\\

$\gamma$ & - & - & $-1.2^{+0.5}_{-0.7}$& $-0.9^{+0.4}_{-0.6}$&$[-3,1]$\\
\hline

$\alpha$ & $2.4^{+0.293}_{-0.264}$ & $2.3^{+0.3}_{-0.2}$ &$2.3^{+0.3}_{-0.3}$&$2.5^{+0.3}_{-0.4}$&$[1,3]$\\

$\beta$ & $0.72^{+0.16}_{-0.18}$ & $0.73^{+0.18}_{-0.19}$ &$0.73^{+0.18}_{-0.19}$&$-0.0^{+0.4}_{-0.5}$&$[-1,1]$\\

$P_b$ & $4.1^{+0.5}_{-0.5}$ & $4.2^{+0.5}_{-0.5}$ &$4.2^{+0.5}_{-0.6}$&$4.4^{+0.6}_{-0.6}$&$[2,10]$\\
\hline

$\alpha_I$ & -&-  &-&$1.9^{+0.2}_{-0.2}$&$[1.5,3]$\\

$\beta_I$ & -& - &-&$0.3^{+0.2}_{-0.3}$&$[-0.5,0.5]$\\

$P_{b,I}$ & -& - &-&$8.8^{+0.9}_{-1.3}$&$[5,10]$\\
\hline
\hline
{AIC} & $2841$&  $ 2802$ &  $2799$& $2801$\\
{relative likelihood} & $\sim 10^{-10}$ & $0.2$ & $1$ & $0.4$\\

\end{tabular}
\caption{The best fit value of parameters for each occurrence rate density model with 
$68\%$ credible interval and the corresponding AICs. Best models are Power-Gaussian as it has lowest AIC. The last row is the likelihoods of other models relative to the  Power-Gaussian one.}
\label{table:parameter}
\end{table*}

Given the observed properties of a planet sample ('PCs'), the posterior probability of a hypothesis ${\bf \theta}$ (with its attendant parameters) can be calculated using the Bayes' theorem as
\begin{equation}
    P(\boldsymbol{\theta} \mid \{PCs\}) = P(\{PCs\} \mid \boldsymbol{\theta}) \pi(\boldsymbol{\theta})\, ,
    \label{post,likelihood}
\end{equation}
where 
$P(\{PCs\} \mid \boldsymbol{\theta})$ is the likelihood and
$\pi(\boldsymbol{\theta})$ the prior. We use exclusively flat priors.

As the planets are discreet points in the period-radius space, we treat them as Poisson point process, following earlier works  \citep[e.g.,][]{Youdin2011,Burke15,Bryson}. Defining the occurrence rate within a bin $\Delta \ln P\, \Delta \ln R_C$ as
\begin{equation}
\Delta f (\boldsymbol{\theta})  = {{d^2 f}\over{d\ln P d\ln R_C}}  \, \Delta \ln P \, \Delta \ln R_C \,
\end{equation}
where 
$\boldsymbol{\theta}$ stands for an ansatz and its attendant parameters,  we can write the total expected number of planet detection as
\begin{eqnarray}
   N_{\rm exp}& = &
   N_* \int \int {{d^2 f}\over{d\ln P d\ln R_C}} \times \eta (P, R_C) \, d\ln  P d\ln R_C \nonumber \\ 
   & = & N_* \sum_{\rm bins} \Delta f(\boldsymbol{\theta}) \, \times \eta(P, R_C)\, ,
   \label{eq:Nexp}
\end{eqnarray} 
where ${\eta}(P,R_c)$ is the averaged detection completeness within the bin (\S  \ref{sec:completeness}). We choose a fine grid, with 20 period bins (1-16 days) and 30 core-radius bins ($0.6 - 2 R_\oplus$).

The likelihood function is then
\begin{equation}
    P(\{PCs\} \mid \boldsymbol{\theta})=\exp(-N_{\rm exp})\,   \prod^{n_p}_{i = 1}
    \Delta f (\boldsymbol{\theta}) \, ,
    \label{likelihood}
\end{equation}
where the product is carried over each detected candidate. 

\begin{figure}
\includegraphics[width=0.45\textwidth,trim=0 0 0 0,clip=]{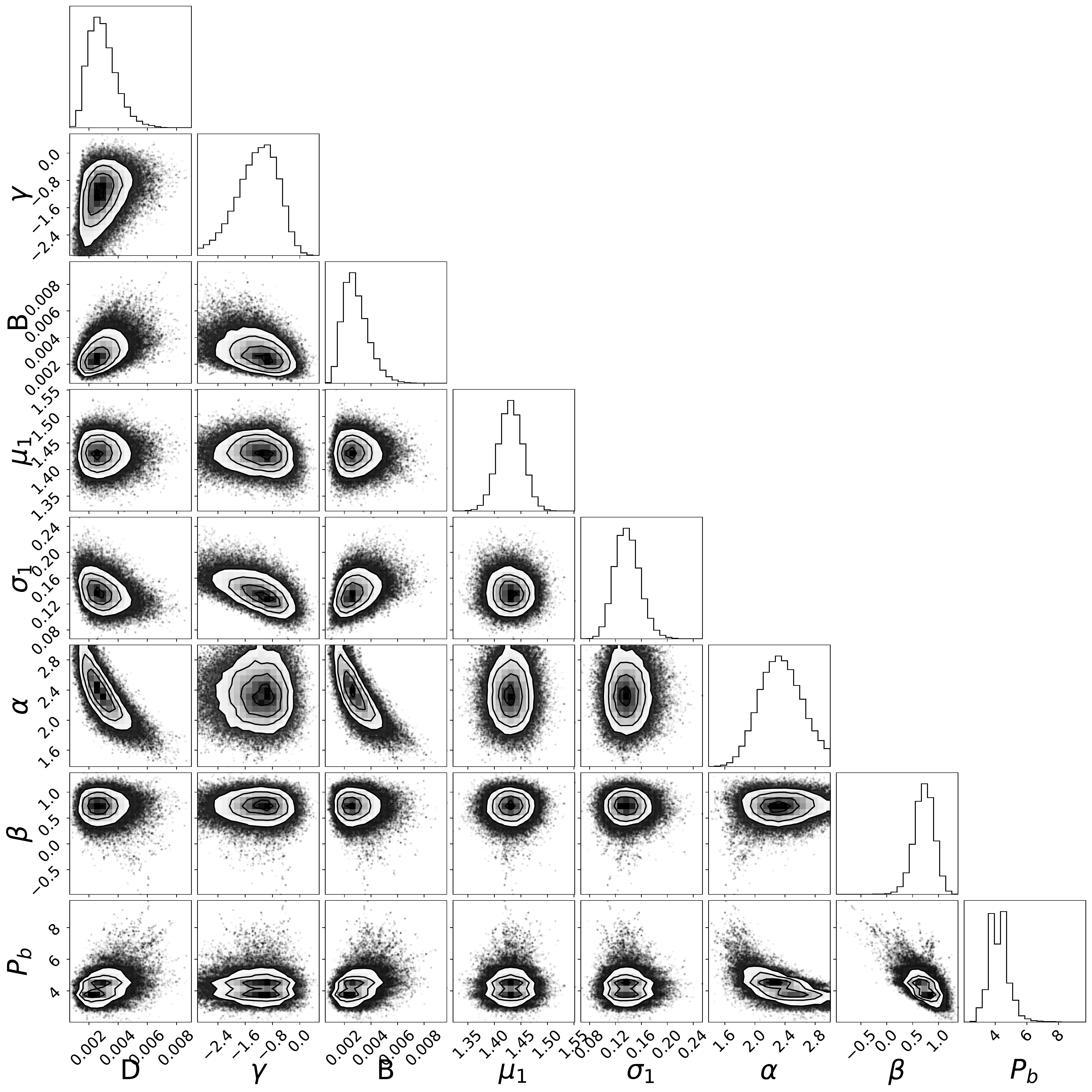}
\caption{Corner-plot for  the Power-Gaussian model, showing the co-variance between  different parameters and their posterior distributions.
}
\label{fig:parameter}
\end{figure}

In detail, we first use
the package \textit{scipy.optimize} to obtain values that maximize the likelihood function. 
We then sample the posteriors using the MCMC algorithm provided by the \textit{emcee} package \citep{emcee}.
We use 50 walkers, 1000 "burn-in" steps and 5000 MCMC steps to find convergence, where the walkers of "burn-in" are initiated in a narrow Gaussian whose mean is the maximum likelihood solution. The parameters corresponding to the median of posterior distribution are shown in Table \ref{table:parameter}, with Fig. \ref{fig:parameter} illustrating the corresponding size distributions for our best models.

To compare the relative merits of various models, we employ the Akaike information criterion(AIC), a measure of information loss. This takes account of both the quality of the fit and the simplicity of the model to avoid both under-fitting and over-fitting. Let the latter be quantified by the number of free parameter,$n_{\boldsymbol{\theta}}$, we write
\begin{equation}
    AIC = 2 n_{\boldsymbol{\theta}}-2
    \ln  \left[P(\{PCs\} \mid \boldsymbol{\theta})\right]\, .
    \label{aic}
\end{equation}
The best model has the lowest AIC score, and the relative merit (likelihood) between models is quantified as $e^{(-\Delta {AIC}/2)}$.
So any model that has an AIC score higher by $\sim 10$ relative to the best model is clearly rejected.

\subsection{Results}

Here, we describe the results of our Bayesian inferences, and compare them against those from IDEM. Their AICs and relative likelihoods are presented in Table \ref{table:parameter}.

First, among the models we tested,  the Bayesian analysis strongly rejects Model 1 (single population) and requires the presence of at least one other population of planets,  in addition  to the super-Earths. Model 1 is extremely unlikely, with a relative likelihood of $\exp(-40/2)\sim 10^{-10}$ compared to the 2-population models.

Second, the 2-period model (Model 4), while allowing for more flexibility in  the period distributions, performs slightly worse than the models that insist  on a single period distribution. In other words, there is no statistical evidence suggesting that the two populations (super-Earths and sub-Earths) have different period distributions, within the limited range that we probe ($1$ to $16$ days). In fact, the obtained period distributions  all feature a power-law rise to a few days, and a roughly logarithmically flat distribution afterwards, analogous to earlier studies using more extended samples \citep{Silburt,Mulders2018,NeilRogers}.

\begin{figure}[h]
\includegraphics[width=1\linewidth]{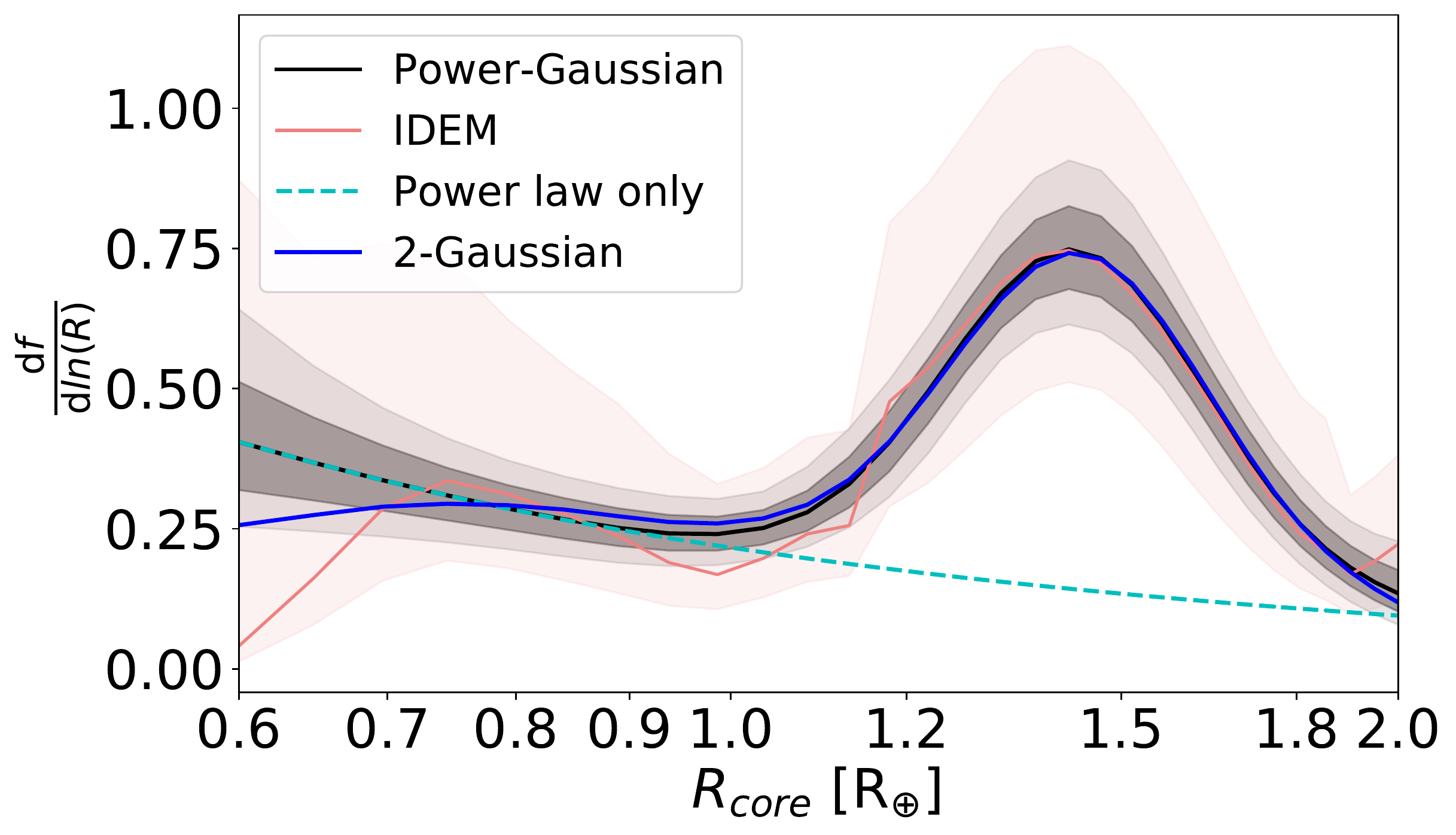}
\caption{Comparison of results from both the Bayesian analysis (Power-Gaussian  model in  black line, dark-grey  and light-grey zones indicating 
$68\%$ and $95\%$ credible intervals,  respectively) and the parameter-free IDEM analysis (red curve, pink shadow zone being $95 \%$ credible interval). The blue curve shows the best-fit 2-Gaussian model. All models agree on the fall of the super-Earths and the emergence of a separate population at smaller sizes.
{In addition, the broke cyan line shows the contribution of the power-law component (sub-Earths) in the Power-Gaussian model.}
}
\label{fig:idembay}
\end{figure}

Third, Models 2 and 3 are statistically indistinguishable (Model 3 is slightly preferred).  Their size and period distributions also look similar (Fig. \ref{fig:idembay}). This is because the Gaussian distribution of the sub-Earths, in the 2-Gaussian model, is very broad and is compatible with a power-law.
From now, on, we choose to focus on Model 3, the so-called power-Gaussian model.

In Model 3, the super-Earths are centered around $1.4 R_\oplus$ with a narrow width of $\sigma_1 \sim 0.13$. In planet mass, these correspond to a centroid of  $\sim 4 M_\oplus$ and a FWHM of a factor of $3.2$. Given that we are focussing on host stars of lower masses (median mass $0.89M_{\odot}$), this centroid falls below that obtained  by    \citet{Wu2019}, $M_p \approx 8 M_\oplus (M_*/M_\odot)$. The reason of the discrepancy is unclear for now. The FWHM does agrees with that in \citet{Wu2019}. The super-Earth population appear to peak narrowly in core mass, a mystery of the formation.

What is of most interest to us is the size distribution of the sub-Earths. The power-Gaussian model yields a size index $\gamma = -1.2 ^{+0.5}_{- 0.7}$, or, the number of sub-Earths most likely rises towards small sizes. This is in drastic contrast to the super-Earths who fall off steeply towards small sizes. In addition, not all objects bigger than  Earth belongs to the super-Earth group. Even at the super-Earth peak ( $\sim 1.4 R_\oplus$), sub-Earths make a non-negligible contribution ($\sim 24\%$). Lastly, when we integrate all planets inward  of $16$ days and with core-sizes from $0.6$ to $2 R_\oplus$, we find that super-Earths and sub-Earths are comparable in  numbers  
($21\pm 4 \%$ vs.  $26\pm 4\%$).

There is  a possibility that the negative $\gamma$ value is driven not by sub-Earths below $1R_\oplus$, but by larger
planets. Following the advice of the reviewer, we experiment with a more sophisticated model that includes a Gaussian (for the super-Earths) and a broken-power law. This introduces two new parameters, the break-radius, and a new power-law index. We find a break-radius of $0.77 R_\oplus$, 
with the power-index above this size to be  $-1.8\pm 0.8$, even steeper than our original result ($-1.2^{+0.5}_{-0.7}$). Due to the lack of objects below this radius,  the index below is very uncertain ($0.7^{+1.0}_{-1.7}$). This allows us to conclude that the rising power-law we find is natural to the data and is not an  artefact of our adopted model.

To compare the parameter-free IDEM results against the Bayesian results,
we first convert the planet observed radii to core radii, as detailed above. We then interpolate the IDEM results between bins, using a third-degree spline smoothing, to  obtain the desired core-size distribution, $df/d\ln R_C$. 
The comparison is shown in Figure \ref{fig:idembay}. They generally agrees with each other. And as expected, our treatment of uncertainties in the IDEM  approach yield larger  error-bars compared to those from forward models.

\section{Discussions}

The decisive evidence for the new sub-Earth population is the  clear inflection point in the size distribution, located around $1R_\oplus$ (Fig. \ref{fig:idembay}).
From the peak of the super-Earths ($1.4 R_\oplus$) to this point, the occurrence rate of super-Earths drops markedly by a factor of a few; while the sub-Earth population   take over gradually in dominance. 
The latter is only revealed thanks to the narrow mass distributions of the former. The mass distribution of the the sub-Earths appear much wider and can be described by a rising power-law towards small sizes. So in our model, this inflection point is actually a 'gap'.

Here, we discuss the robustness of these results and comparing them against previous works.

\subsection{Sample Cuts and  Other variations}
\label{subsec:cuts}

Our analysis is performed on a small set  of 'clean' planet sample ($298$ in total). Here, we vary the sample criteria and observe how the results  for the power-gaussian model change. The priors remain the same as before. 
For each experiment, we  remove (or add) one constraint on the property of the stars or PCS, keeping other cuts unchanged. 

\begin{figure}
    \centering
    \includegraphics[width=0.48\textwidth]{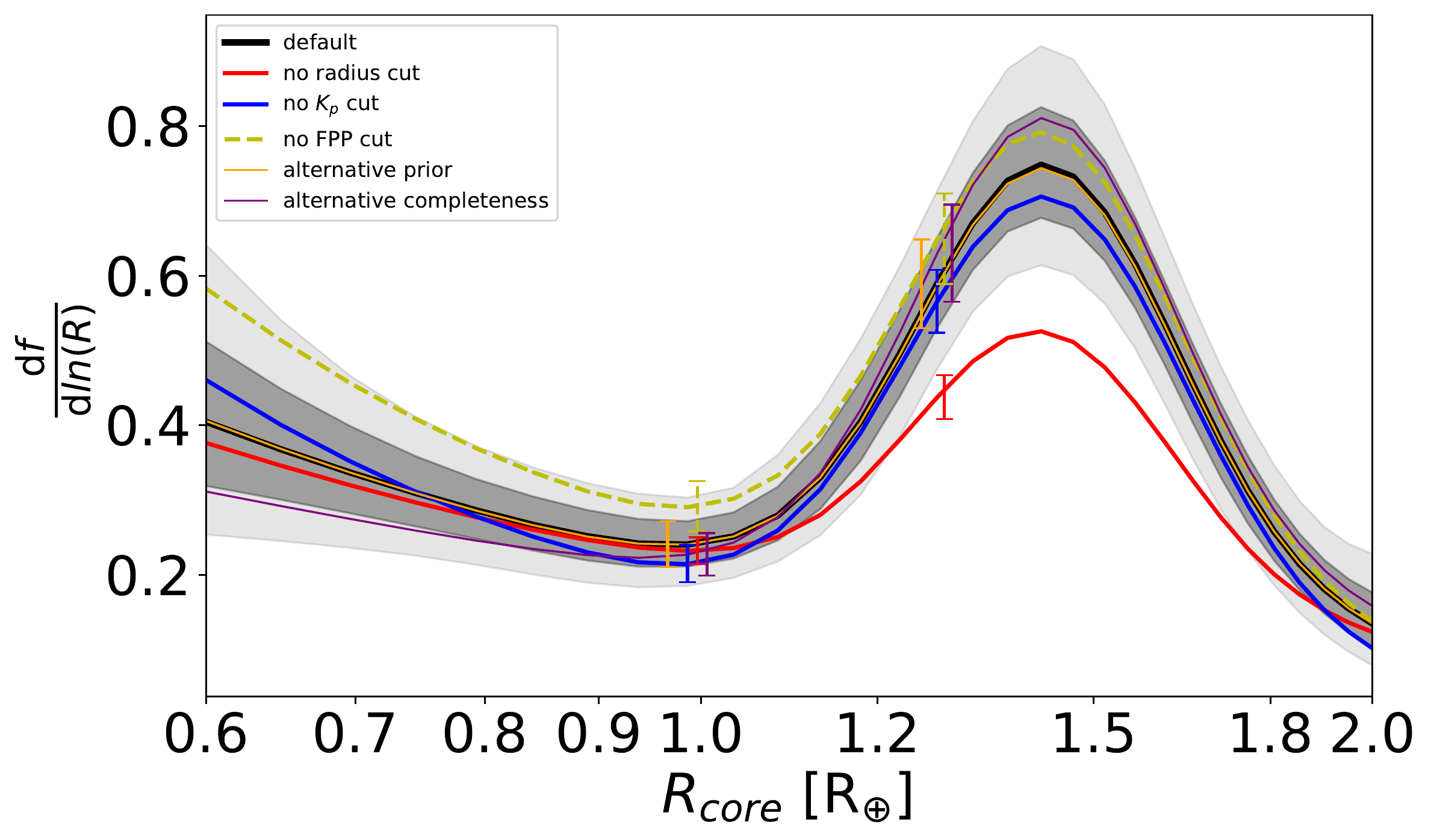}
    \caption{Comparing the core-radius distribution across models. The black curve and the shaded belts are as those in  Fig. \ref{fig:idembay}, while the colored curves from different experiments, and the error-bars marking  their respective  
    $68\%$ credible interval (only  at two select radii for clarity).
    Regardless of sample selection or other variations, super-Earths are found to have a narrow size distribution. With the exception of the 'confirmed' cut, a new component (sub-Earth) that rises towards small sizes is required in all models.
   }
    \label{fig:cutcompare}
\end{figure}

We also experiment with alternative priors and completeness treatment. 
All experiments yield similar results compared to our default case (see Fig. \ref{fig:cutcompare}). 
We comment on each briefly below.

\begin{itemize}

\item No stellar radius cut: we experiment with employing all stars, regardless of radii. We exclude stars that satisfy the following crude cuts, 
\begin{equation}
    \left\{
    \begin{array}{@{}lr@{}}
    \frac{R_*}{M_*^{0.9}} > 1.3, & \text{if}\    M_* <M_{\odot} \\
\frac{R_*}{M_*^{1.25}} > 1.3, & \text{otherwise}
\end{array}\right.
        \label{eq:evoflag}
\end{equation}
as they are most likely evolved. 
There are then 
$48642$ stars and 
$794$ PCs. The stellar sample has expanded by a factor of 
3.7, while the planet sample only a factor of 
$2.7$, leading to a slightly lower occurrence (Fig. \ref{fig:cutcompare}). Moreover, the super-Earth peak is broadened, due to the known correlation between planet mass and stellar mass. The sub-Earth slope is $\gamma \approx -1.1$.

\item No $K_p$  cut: while dimmer stars return fewer super-Earths and sub-Earths, the completeness correction largely accommodates this correctly. There is no perceptible change to the occurrence rates.

\item No FPP cut:  retaining all objects,  regardless of their FPP values, adds 36 new candidates to our  list, but produces no change to the  results. 
The newly added candidates  do not make the gap features more conspicuous, presumably because they are more likely to be potential false positives.

However, we believe that unconfirmed PCs with a low FPP value are reliable enough and the sub-Earth distribution in our default case is likely genuine.

\item Alternative Priors:
We have chosen broad and flat priors previously. Here, we follow the insights from \citet{Wu2019} and use a narrower prior on the super-Earths: 
$\mu_1$ is now normally distributed with a mean at $1.43 R_\oplus$ and a variance of $0.05 R_\oplus$, and $\sigma_1$ is normally distributed with a  mean  of $0.1$ and a variance of $0.1$. The results remain the same.

\item Alternative completeness: while we have  adopted the \citet{burkeFLIT} to to  obtain  the detection completeness, there exists an alternative approach  \citep{christainsenpl}. In this method, the 
completeness of a detection is expressed as a function of its MES  (multiple-event-statistics) value, as well as some parameters $a,b,c$ that are empirically obtained from pixel-level injection experiments,
\begin{equation}
    \eta(MES,a,b,c) = \frac{c}{b^a\Gamma(a)}\int_0^{MES}t^{a-1}e^{-\frac{t}{b}} dt\, .
    \label{gamma}
\end{equation}
Because pixel-level injections are computationally expensive, typically one injection per star is possible. These empirical parameters are therefore best fits for the entire Kepler stellar sample. This type of correction is used by e.g., \citet{NeilRogers,hsu2019}. Here, we employ   KeplerPORTS to map radius and period into MES for every planet host in our sample, and then adopt $a=
29.41$, $b=0.284$, $c=0.891$ from \citet{NeilRogers}. 
Running the Bayesian analysis return results that do not differ significantly from our original ones (see Fig. \ref{fig:cutcompare}).

\item Reliability correction: as some planet candidates are less reliable (a higher false positive probability) than $100\%$, we experiment with down-weighting them in the analysis and repeat the IDEM analysis. This makes negligible difference, presumably because all candidates are above $90\%$ reliable, with most close to $100\%$.

\end{itemize}

\subsection{Comparison with Other works}

\begin{figure}
    \centering
    \includegraphics[width=1\linewidth]{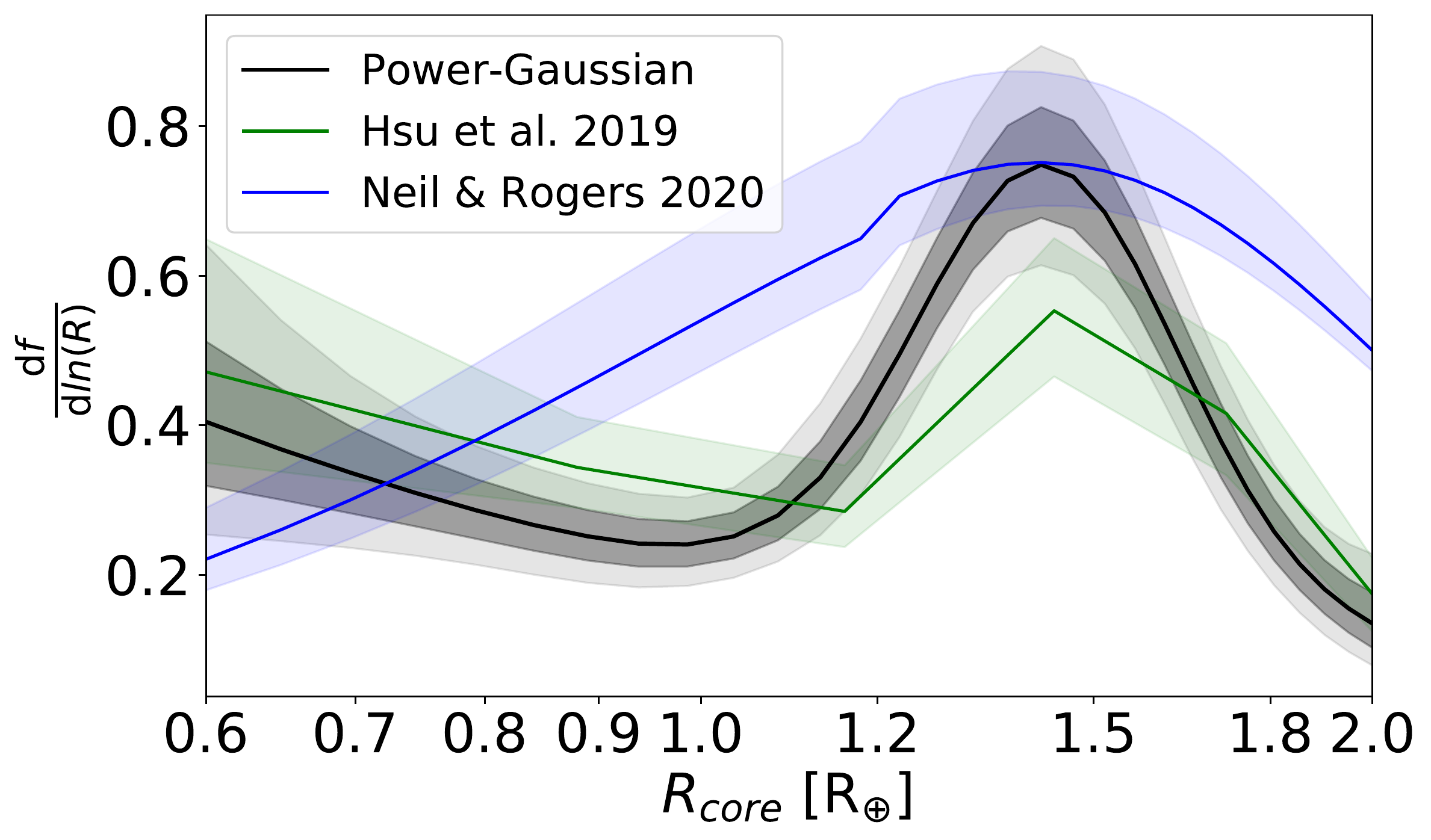}
    \caption{Comparison of our core-radius distribution (same model as that in Fig. \ref{fig:idembay}) and those from \citet{hsu2019} (green) and \citet{NeilRogers} (blue), for planets with periods from $1$ to $16$ days. The color shaded regions indicate their respective $68\%$ credible interval. Three studies differ in  sample selection, completeness correction and inversion  method.  }
    \label{fig:neilhsu}
\end{figure}

There are two recent works that employ the updated completeness corrections and have included sub-Earths into their modelling efforts \citep{hsu2019,NeilRogers}. We compare our results with theirs here, for the period range from $1$ to $16$ days.. The \citet{NeilRogers} work infers distributions in planet masses. We  use simple scaling relations of $M/M_\oplus = (R_C/R_\oplus)^4$ for $R_C >1.2 R_\oplus$ and $M/M_\oplus = (R_C/R_\oplus)^3$ for $R_C <=1.2 R_\oplus$ to translate their mass distribution into core radius distribution. 
The other study \citep{hsu2019} infers the distribution in planet transit radius. 
We convert from these to core-radius as laid out in \S  \ref{sec:models} and interpolate between bins where  necessary. The comparisons are displayed in  Fig. \ref{fig:neilhsu} and we briefly describe any similarities or differences here.

Similar to our study here, both works show a  droping-off of super-Earths towards small sizes, and a new population below about $1 R_\oplus$. 

In the parameter-free, grid-based inversion model of \citet{hsu2019}, one observes a radius gap between super-Earths (including Neptunes) and sub-Earths.
However, because their stellar sample covers a wide range of spectral types (mostly FGK dwarfs with no cuts in stellar radius), their super-Earth peak is broader and lower in amplitude, resembling closely our 'no-radius-cut' case in Fig. \ref{fig:cutcompare}. By focussing on a narrow range of spectral type, we end up with a smaller planet sample, but can detect more prominent features.
 
 The forward-modelling approach of \citet{NeilRogers} is very similar in spirit to our Bayesian models. By parametrizing the size distributions as a number of distinct Gaussians, they find that a new population of small planets (which they term 'intrinsically rocky') is strongly demanded by data.
However, their overall size distribution does not show a clear inflection point, let alone a 'gap' as we find here. We compare their results with our 2-Gaussian model in detail. Their best-fit super-Earth peak lies at $R_C = 1.53 R_\oplus$ (slightly larger than our $1.44 R_\oplus$), but with a much broader dispersion of $\sigma = 0.63$ (compared to our $0.14$ $\pm 0.03$). We attribute this latter difference to their wide range of stellar properties. They include almost all dwarfs, and since super-Earth properties are known to correlate with stellar properties, this tends to broaden the super-Earth peak. More significantly, their sub-Earths are described by a log-normal distribution with a peak at size $R_C \approx 0.96 R_\oplus$ (compared to our $0.74$), and a radius dispersion of $\sigma \sim 0.4$ (same as our $0.4$). The fact that their sub-Earths are more massive than ours, and that their super-Earths are more broadly distributed, give rise to a gap-free transition from  the super-Earths to the sub-Earths (Fig. \ref{fig:neilhsu}), as well as a dropping occurrence towards very small sizes.
These are qualitatively different from our results and those in  \citet{hsu2019}. We believe that our choices of parameter space (narrower while preserving almost all sub-Earth candidates) is  better suited to resolve any possible feature in the size distribution. 
A recent investigation, using their original model but with more relaxed assumptions, leads to better agreements with our results (Neil, private communication).

\subsection{Insights from Ultra-Short-Period Planets}
\begin{figure}[b]
\includegraphics[width=1\linewidth,]{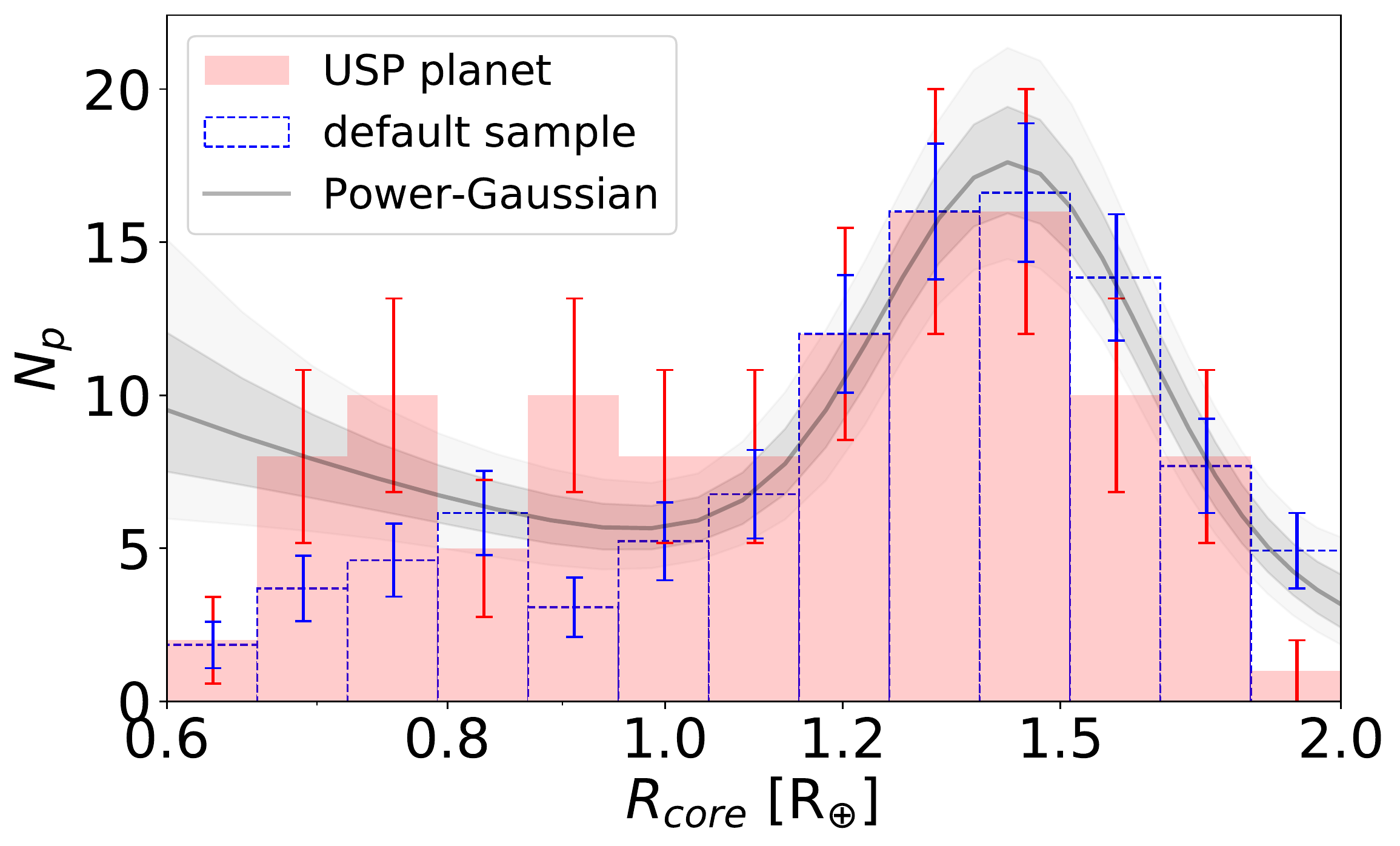}
\caption{Comparing the observed size distributions of ultra-short-period planets ($<1$ day) (red histogram, uncorrected for completeness) with our default sample (blue histogram, $1-16$ days , uncorrected), and our power-Gaussian model (gray curve, corrected). The latter two are scaled to have the same heights as the USPs at $1.3 R_\oplus$. The USPs show an excess of sub-Earths than our planet sample, due to a better completeness. They therefore provide support to our model.}
\label{fig:usp}
\end{figure}

We have selected an  optimum range to study the sub-Earths. Our choices of $1$ to $16$ days and bright hosts,  minimize the uncertainties due to extreme completeness correction, without sacrificing many candidates. We have high confidence in our results because the corrected model actually does not look dramatically different from the raw data (Fig. \ref{fig:usp}).

Another support for our model comes from an intriguing  class of objects, the so-called ultra-short-period planets (USP for short), planets that have orbital periods short-ward of one day \citep[e.g.][]{Rappaport,SO2014,Winn2018}. 
At these short periods, {\it Kepler} can  be sensitive to much smaller planets than otherwise possible. The origin for these  boiling worlds is still under debate \citep[e.g.][]{Lee2017,Petrovich2019,Pu2019,Millholland2020}, though scenario involving  dynamical migration by companion planets have received some observational supports  \citep{SO2014,Adams2020}.  If so, their mass distribution should be similar to those further-away. Here, we exploit this advantage to shed light on our model.\footnote{There is another advantage to using USPs. Ultra-short-period candidates are less likely to be false positives due to background eclipsing binaries. At such short periods, ellipsoidal variations would have been clearly visible.}

Fig. \ref{fig:usp} compares the observed size distribution of USPs ($< 1$ day), against that of our default sample ($1-16$ days), and against our preferred model. These $127$ USPs are discovered by the {\it Kepler} main mission, are smaller than  $2 R_\oplus$, include both "confirmed" and "candidate" dispositions. Their host magnitudes are mostly  between $14$ and $16$ mags. We have also updated their radii using GAIA stellar parameters \citep{Berger2020star}.

Compared to our planet sample, the USPs show an excess of sub-Earths, thanks to their better completeness in these small sizes. 
More importantly, one can  see a near perfect agreement between the  un-corrected USPs and our power-Gaussian model.
If anything, the USPs suggest that the sub-Earth rise may be steeper than our current model, after 
completeness correction is properly applied.

Other than supporting our model, this result also suggests that USPs are indeed formed similarly as planets further away. 

\subsection{Siblings of the sub-Earths}
\begin{figure}
\includegraphics[width=1\linewidth]{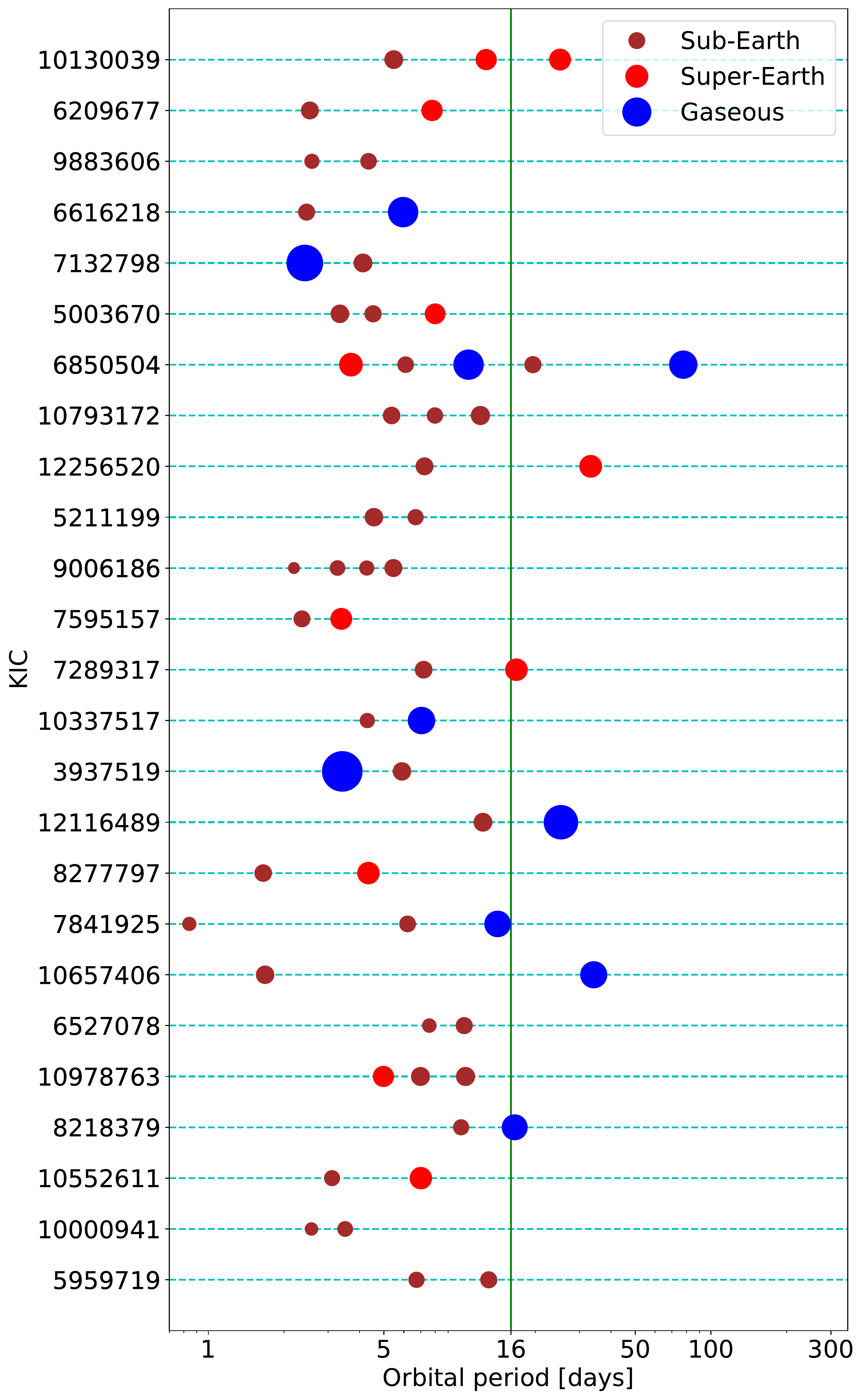}
\caption{About half sub-Earths in our default  sample have  transiting  planet siblings. They are plotted here according to  their sizes. 
The systems are arranged  from top to  bottom by decreasing host star masses. 
}
\label{fig:sibling}
\end{figure}
In our sample, there are 71 PCs with radii smaller than $1.1R_{\oplus}$. These  most likely belong to the sub-Earth population, according to our Power-Gaussian model. Among them, 33 sub-Earths have at least one (though frequently more than one) transiting sibling planet. These multiple-planet systems are shown in Figure \ref{fig:sibling}. We see that sub-Earths appear to happily co-exist with larger planets (super-Earths, mini-Neptunes, even Jovians), with no substantial differences in both their period distribution and their stellar mass distribution. 

The fact that about half of all sub-Earths have transiting siblings may indicate that the majority of them prefer to reside in  systems with other larger planets.

\section{Conclusion}

The primary goal of the {\it Kepler} mission is to find Earth analogs in the  habitable zones \citep{Koch2010}. It has had some successes in finding rocky planets around low
mass stars where the habitable zones are close-in \citep[see e.g.,][]{Borucki2013,Kane2016}, but there are few solid such detections around Sun-like stars \citep[see Fig. 3 in][]{Borucki2019}. While {\it Kepler} did find an abundance of super-Earths, these are likely stripped-down version of the mini-Neptunes and do not exist beyond a few tens of days.

In this study, by choosing a mostly complete planet sample, by assuming simple models with broad priors, we are able to establish that the core-radii of super-Earths peak narrowly around $1.4 R_\oplus$ (corresponding to $\sim 4 M_\oplus$ for terrestrial composition). At smaller sizes, we find that these planets disappear rapidly in number, with the data showing a clear inflection point, and most likely a gap, between super-Earths and a new population of small planets (sub-Earths).  The existence of an inflection point is robust, regardless of sample selection, inversion method, or completeness correction. 
The size distribution of ultra-short-period planets provides additional supports to our model.

We find that the sub-Earths, dominant in number below $\sim 1 R_\oplus$,  have a similar period distribution as the super-Earths, co-exist with larger planets in the same planetary systems, and, most interestingly, can be described by a rising power-law towards small sizes. This is in contrast to the nearly singular distribution of the super-Earths.

 This implies that sub-Earths and super-Earths are formed differently. While super-Earths (and their cousins that still retain the primordial H/He envelopes, mini-Neptunes) are most likely formed in the gaesous proto-planetary disks (and therefore may be called Generation-I planets), the sub-Earths, we speculate, may form later in a collisional debris disk, much like the terrestrial planets are theorized to do \citep[e.g.][]{Chambers1998,Kokubo,Morishima2008}. In such a debris disk, the final planetary outcome depends on the initial mass budget. One therefore expects the planetary masses to be broadly distributed.
 We argue that  these planets may be thought of as Generation-II planets.

If our speculation, that these planets are true terrestrial analogs, is correct, we expect them to extend to at least the AU region. They are then  likely the only population of rocky planets in the habitable zones, and they are the most likely habourer of life. While  {\it Kepler} just fell short of discovering them in AU distances, more sensitive new missions  should be organized to go after these targets.

To provide a rough guidance for these future searches, we boldly estimate the occurrence rate of sub-Earths in the habitable zone (Table \ref{table:eta}),  extrapolating from models that  apply only  within $16$ days.
The estimates show substantial  uncertainties, largely because of our short lever arm in constraining the period distribution. Even an extension by a factor of $2$ in period would have helped firming down the $\eta_\oplus$ value substantially.

\begin{table}[h!]
\centering
\begin{tabular}{|c|c|c|}
\hline
Models & $f$ (1-16 days) & $f$ (200-400 days, $\eta_\oplus$)\\
    \hline 
Power-Gaussian & $26^{+4}_{-4}\%$ &  $143^{+138}_{-73}\%$  \\
2-Gaussian &     $24^{+6}_{-7}\%$ &  $129^{+135}_{-67}\%$\\
2-period &  $23^{+4}_{-4}\%$    & $8^{+30}_{-7}\%$  \\
\hline
\end{tabular}
\caption{Estimated occurrences of sub-Earths ($0.6-2R_\oplus$) around GK dwarfs.}
\label{table:eta}
\end{table}

\acknowledgements

This work makes use of the NASA Exoplanet archive, and results from the NASA Kepler mission, the ESA Gaia mission and the California-Kepler survey.  Danley Hsu, Andrew Neil, Eric Ford and an anonymous referee provided helpful feedbacks.
We acknowledge NSERC for a research grant, including support for QYS as a summer undergraduate researcher. 

\software{Robovetter (Thompson et al. 2018; Coughlin 2017), KeplerPorts \citep{Burkemodel}, emcee \citep{emcee}, corner \citep{corner}, Scipy \citep{scipy}, Numpy \citep{numpy}, Matplotlib \citep{matplotlib}, pickle \citep{pickle}}

\bibliography{ref}

\end{CJK*}

\end{document}